\documentclass[aps,superscriptaddress,reprint,longbibliography,onecolumn]{revtex4-2}
\usepackage{amsmath}
\usepackage{graphicx,textcomp}
\usepackage{enumerate,subeqnarray}
\usepackage{verbatim,natbib}
\usepackage{amsmath, amsthm, amssymb}
\usepackage{amsfonts}
\usepackage[colorlinks=true, citecolor=red, linkcolor=blue, urlcolor=blue ]{hyperref}

\newcommand{\oo}[1]{\mathcal O}

\usepackage{orcidlink}
\begin{document}
	
	\title{Analytical methods for cytoplasmic streaming in elongated cells}
	
	\author{Pyae Hein Htet\,\orcidlink{0000-0001-5068-9828}}
	\affiliation{Department of Applied Mathematics and Theoretical Physics, University of Cambridge,	
		Cambridge CB3 0WA, UK}
	\author{Eric Lauga\,\orcidlink{0000-0002-8916-2545}}
	\email{e.lauga@damtp.cam.ac.uk}
	\affiliation{Department of Applied Mathematics and Theoretical Physics, University of Cambridge,	
		Cambridge CB3 0WA, UK}
	
	\date{\today}

	\begin{abstract}
		
		Cytoplasmic streaming, the coherent flow of cytoplasm, plays a critical role in transport and mixing over large scales in eukaryotic cells. In many large cells, this process is driven by active forces at the cell boundary, such as cortical cytoskeletal contractions in \textit{Drosophila} and \textit{C.~elegans} embryos, or intracellular cargo  {transport} in plant cells. These cytoplasmic flows are  approximately Newtonian and  governed by the Stokes equations. In this paper, we {use} lubrication theory --  {a powerful technique for simplifying the fluid mechanics equations in elongated geometries} -- to derive a general solution for boundary-driven cytoplasmic flows.   We apply this framework to predict cytoplasmic fluid dynamics and cortical stresses in four  {systems of biological significance}: the \textit{Drosophila} and \textit{C.~elegans} embryos (including pseudocleavage furrow formation),   the pollen tube of seed plants, and plant root hair cells.  Our results showcase the elegance and accuracy of   asymptotic solutions  in capturing the complex flows and stress patterns in diverse biological contexts, reinforcing its utility as a robust tool for cellular biophysics.

	\end{abstract}
	
	\maketitle
	\clearpage
	
	\section*{Significance statement}
	Cytoplasmic streaming, the directed, coherent flow of cytoplasm within cells, enables vital transport and mixing in large biological cells. In many elongated cells, active forces at the cell boundary drive this flow, and the governing fluid dynamics equations simplify, permitting analytical solutions. By applying this approach, we characterize here the intricate flow patterns and stress distributions across four distinct biological systems: fruit fly and roundworm embryos, pollen tube cells, and root hair cells. 
	{Our results reveal the accuracy and utility of fluid mechanics  as a versatile tool in cellular biophysics and point to its potential for explaining complex cellular transport phenomena.} This work strengthens the role of theoretical modeling in advancing our understanding of active intracellular processes.

	\section{Introduction}
	
	{Cytoplasmic streaming is the coherent bulk flow of the cytoplasm, commonly seen in large eukaryotic cells. In these cells, passive diffusion is too slow to effectively transport cellular constituents across long length scales, and streaming provides an active advection-based transport mechanism to sustain the cell's metabolic and developmental needs~\cite{goldstein2015}}. Since the discovery of cytoplasmic streaming in the 18$^{\text{th}}$ century in the green algae \textit{Nitella} and \textit{Chara}, cytoplasmic flows in plants have been extensively studied in the biology and biophysics literature~\cite{allen1978, kamiya1981, shimmen2004}. In large plant cells, cargoes such as vesicles, chloroplasts, and other organelles are actively dragged by molecular motors along intracellular filaments;  {these filaments are typically bundles of actin, but microtubules-based transport has also been documented~\cite{romagnoli2003}}. This forcing induces coherent flows of the cytoplasm, which further facilitates transport not only by direct advection, but also by enhanced diffusivity via Taylor-Aris dispersion~\cite{kikuchi2015}, illustrating the diverse physical and fluid dynamical effects at play inside biological cells.
	
	Although most common in plants and algae, cytoplasmic streaming occurs also in various other cells, including amoebae~\cite{stockem1983}, fungi~\cite{pieuchot2015}, slime moulds~\cite{pringle2013}, and animal oocytes and embryos~\cite{quinlan2016, lu2022}. In animal embryos, cytoplasmic flows are often driven by the cell cortex, a cytoskeletal actomyosin layer at the cell boundary which is flexible enough to flow as a fluid and stiff enough to enable the cell's mechanical functions~\cite{salbreux2017, chugh2018}. Cortex-driven cytoplasmic flows play key roles in early animal development, for instance, in the segregation of yolk granules from the rest of the cytoplasm in zebrafish zygotes~\cite{shamipour2019, fuentes2010}. Homogeneous  distribution of the numerous nuclei obtained via multiple rounds of {nuclear} division in the \textit{Drosophila} embryo~\cite{deneke2019, lopez2023} and cell polarization via asymmetric transport of PAR proteins  {(a group of polarity-regulating proteins)} and other cellular constituents~\cite{gubieda2020}  {in the \textit{C.~elegans} embryo} both rely on cortical and cytoplasmic flows. 
	
	The cytoplasm is a complex and crowded medium, but its behavior is well approximated as an effective Newtonian fluid over the length and time scales at which streaming is observed, a well-established modeling assumption supported by rheological data~\cite{ganguly2012} as well as particle image velocimetry (PIV) and computational studies~\cite{deneke2019, klughammer2018, niwayama2011}. This makes cytoplasmic flows an ideal subject for applying the wealth of fluid dynamical research developed over centuries.
	
	However, even with a Newtonian approximation for the bulk cytoplasmic fluid, the intricate biomechanical coupling between the bulk fluid and the cell boundary often requires complicated computational approaches to model in detail~\cite{bacher2019, bacher2021, lopez2023}. Nonetheless, simplified models of the active boundary forcing as a boundary slip velocity capture the large-scale behavior of the bulk cytoplasmic flow and provide valuable biological insight~\cite{niwayama2011, deneke2019, htet2024}. When, in addition, the flows are axisymmetric and the geometry is sufficiently elongated, the analysis is greatly simplified by a long-wavelength approximation, known in the fluid dynamics community as ``lubrication theory"~\cite{langlois1964,leal2007,ockendon1995}, resulting in fully analytical solutions.
	
	Although its beginnings and namesake lie in the thin lubricant layers in fluid bearings~\cite{reynolds1886, hamrock2004}, lubrication theory pertains more generally to flows in which one length scale is much smaller than the others. It is an approximation to the full Navier-Stokes equations for Newtonian fluid flow which exploits this disparity in length scales. Beyond industrial and engineering applications~\cite{lugt2011}, lubrication theory informs fundamental fluid dynamical phenomena such as viscous gravity currents, thin liquid films, and flows through narrow gaps~\cite{greenspan1978,smith1973,huppert1982,oron1997,huppert2006,stone2005}, and has applications in areas as diverse as geophysics (e.g.~lava domes~\cite{huppert2006}, faults~\cite{brodsky2001}, ice sheets~\cite{schoof2013,robison2010}) and biomedicine (e.g.~fluid dynamics of the eye~\cite{siggers2012,fitt2006}, cartilage and joint lubrication~\cite{burris2017}). 
	
	Despite the wide applicability of lubrication theory, studies which exploit this powerful tool in the context of cytoplasmic streaming  {are few and far between}. In this paper, we show how lubrication theory can be used to characterize axisymmetric cytoplasmic flows, and the resultant stresses, in elongated cells; we thus   present analytical expressions for the cytoplasmic flow field and cortical stresses, bypassing the need for intensive numerical computations.  {Using} {as model systems} (i) the \textit{Drosophila} embryo, (ii) the \textit{C.~elegans} embryo, (iii) the pollen tube of flowering plants, and {(iv) plant root hair cells}, we demonstrate the versatility and accuracy of our approach in characterizing flows and stresses in a range of biological cells.  {The centerline cytoplasmic flows in \textit{Drosophila} embryos are directed from the center towards the poles, thus facilitating a symmetric distribution of nuclei. In contrast, the unidirectional centerline cytoplasmic flows in the \textit{C.~elegans} embryo contributes to the breaking of symmetry. The pollen tube features a distinct geometry, and in the root hair cells we place a particular focus on helical streaming. Therefore, the four problems we examine represent not only biologically, but also fluid mechanically, distinct systems.} Our methods are  simple, and yet  {broadly applicable and} model cytoplasmic flows {with great accuracy}.

	In \S\ref{sec:maths}, we present our general mathematical model and derive analytical expressions for the cytoplasmic flow field, {streamlines} and cortical stresses inside an elongated cell in terms of the prescribed boundary forcing. In subsequent sections, we focus on applications, first addressing cytoplasmic streaming in the \textit{Drosophila} embryo in \S\ref{sec:drosophila}. We then characterize flows and stresses inside the \textit{C.~elegans} embryo in \S\ref{sec:celegans}, reproducing the results of a numerical study~\cite{niwayama2016} with our analytical model, and further model cytoplasmic streaming during pseudocleavage. In \S\ref{sec:pollentubes}, we extend our methods to an annular geometry, and characterize streaming in pollen tubes while  {\S\ref{sec:roothaircells} further generalizes our model to incorporate axisymmetric azimuthal flows and describes helical streaming in plant root hair cells.} We end with a discussion in \S\ref{sec:discussion} of the merits of our model, its limitations, and potential generalizations. 
	
	\section{Solution for boundary-driven flows in an elongated cell}\label{sec:maths}

	\begin{figure}[t!]
		\includegraphics[width=0.6\textwidth]{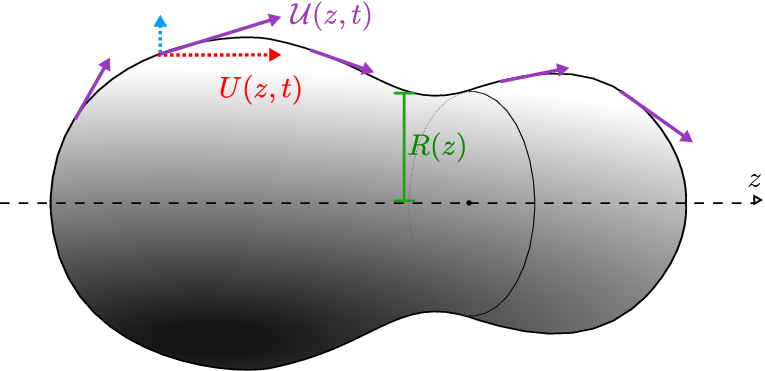}
		\caption{Cytoplasmic flows inside an elongated cell with long axis in the $z$ direction and cylindrical radius $R(z)$. Intracellular flows are driven by a slip velocity $\mathcal{U}(z,t)$ tangential to the cell boundary (schematized as purple arrows) which has an axial component $U(z,t)$ (schematized as red arrow).}\label{fig:setup}
	\end{figure}
	
	\subsection{{Setup} and governing equations}
	
	Working in cylindrical coordinates $(r,z)$, we model an elongated cell as an axisymmetric domain with a rigid boundary given by $r = R(z)$; see Fig.~\ref{fig:setup}. The inside of the cell is filled with cytoplasm modelled as an incompressible Newtonian fluid of dynamic viscosity $\mu$ and density $\rho$. A cytoplasmic flow $\mathbf u(r,z,t)$ is driven by an active forcing at the boundary in the form of a prescribed axisymmetric slip velocity $\mathcal U(z,t)$ tangential to the cell boundary. 
	
	Since most biological cells are small, the relevant velocity scale $V$ and length scale $R$ are sufficiently small that inertia may be neglected. More specifically, the Reynolds number $Re := \rho VR/\mu$, a dimensionless number which characterizes the importance of inertial effects relative to viscous effects, is close to zero. {In the four biological systems we will consider, assuming a kinematic viscosity close to that of water, $\mu/\rho \approx 10^{-6} \text{m}^2$/s (a lower bound),  the Reynolds number ranges from below $10^{-6}$ in the 
	\textit{C.~elegans} embryo ($U \sim 0.1$ \textmu{m}/s, $R \sim 10$ \textmu{m}) to below $10^{-4}$ in the \textit{Hydrocharis} root hair cell ($U \sim 5$ \textmu{m}/s, $R \sim 20$ \textmu{m}).} The mathematical consequence of a negligible Reynolds number is that the nonlinear, inertial terms in the Navier-Stokes equations which govern the motion of an incompressible Newtonian fluid may be neglected. The equation of motion of the cytoplasmic flows thus reduces to the incompressible Stokes equations
	\begin{equation}\label{eq:stokes}
		\mu\nabla^2 \mathbf u = \nabla p, \qquad \nabla \cdot \mathbf u = 0,
	\end{equation}
	subject to the boundary condition 
	\begin{equation}\label{eq:stokesbc}
		\mathbf u = \mathcal U(z,t)\mathbf t(z),
	\end{equation} 
	at the surface $r = R(z)$ of the cell. Here $p$ is the pressure inside the cell  {and $\mathbf t(z)$ is the unit tangent vector pointing in the direction of increasing $z$.}

	\subsection{Lubrication  {approximation}}	
	{We will now  exploit the elongated geometry of the cell and simplify the system \eqref{eq:stokes} and \eqref{eq:stokesbc} using lubrication theory.  Formally, two conditions are necessary for the lubrication approximation to be justified quantitatively~\cite{leal2007}. The first condition is that the longitudinal length scale $L$ is much larger than the radial length scale $R$, which implies the intuitive picture of any small cross-section of the cell looking like part of a locally-straight cylindrical pipe. Although this is a formal mathematical requirement, we will see below that lubrication theory actually performs surprisingly well at moderate elongation. The second condition is that the reduced Reynolds number, $Re_r := ReR/L$, is small, allowing us to neglect inertia; this holds trivially since the Reynolds number $Re$ is already small in all the micro-scale biological systems considered below.}
	
	{Under the lubrication approximation, the radial component of the Stokes equations simplifies to}
	\begin{equation}
		\frac{\partial p}{\partial r} = 0,
	\end{equation}
	i.e.~there are no radial pressure gradients to leading order.  {Writing $\mathbf u(r,z,t) = u(r,z,t) \mathbf e_z + v(r,z,t) \mathbf e_r$,} the longitudinal component reduces to 
	\begin{equation}\label{eq:stokeslong}
	\frac{\mu}{r}\frac{\partial}{\partial r}\left( r\frac{\partial u}{\partial r}\right) = \frac{\partial p}{\partial z}.
\end{equation}
	This is therefore pressure-driven pipe flow in a locally cylindrical cell geometry. An {axisymmetric} slip velocity $\mathcal U(z,t)$ {with no azimuthal component} is imposed tangentially to the wall, and the longitudinal flow $u$ satisfies the longitudinal component of this boundary condition, 
	\begin{equation}\label{eq:BC1}
		u(R(z),z,t) = U(z,t),
	\end{equation}
	where, crucially, $U(z,t) := \mathcal U(z,t) \mathbf t(z)\cdot \mathbf e_z$ is the $z$-component of $\mathcal U \mathbf t(z)$. This relation between $U$ and $\mathcal U$ may be expressed more explicitly in terms of $R$ as $U(z,t) = \mathcal U(z,t)/\sqrt{1 + R'(z)^2}$, where $'$ is used to denote a derivative with respect to $z$. Furthermore, mass conservation in a closed cell implies that the volume flux through any cross-section is zero,
	\begin{equation}\label{eq:BC2}
		\int_0^{R(z)} u(r,z,t)r\,\mathrm dr = 0.
	\end{equation}

	\subsection{Velocity field}
	
	The general solution to Eq.~\eqref{eq:stokeslong} is 
	$u = \frac{1}{4\mu}\frac{\partial p}{\partial z}r^2 + A \ln r + B$, where $A$ and $B$ are integration constants. Regularity at the origin requires $A = 0$, while the unknown pressure gradient $\partial p/\partial z$ and the constant $B$ are determined by imposing the conditions \eqref{eq:BC1} and \eqref{eq:BC2}. This 
	yields our solution for the longitudinal flow,
	\begin{equation}\label{eq:u}
		u(r,z,t) = {U(z,t)}\left[2\left(\frac{r}{R(z)}\right)^2 - 1\right].
	\end{equation}
	
	The radial velocity $v$ is then determined by integrating the incompressibility condition, written out in full as
	\begin{equation}
		\frac{1}{r}\frac{\partial(rv)}{\partial r} + \frac{\partial u}{\partial z} = 0,
	\end{equation}
	and using the solution \eqref{eq:u} for $u$ we have just determined. The integration constant vanishes by regularity at $r = 0$, yielding the transverse component of the flow as
	\begin{equation}\label{eq:v}
		v(r,z,t) = \frac{1}{2}\frac{\partial U(z,t)}{\partial  z} r\left[1 - \left( \frac{r}{R(z)}\right)^2 \right] +U(z,t)\frac{\mathrm dR(z)}{\mathrm dz}\left( \frac{r}{R(z)}\right)^3.
	\end{equation}
	We check that $v = R'U$ at $r = R$, and $v$ therefore satisfies (to leading order {in $R'$}) the radial component of the slip velocity condition. We have thus derived analytical expressions, equations \eqref{eq:u} and \eqref{eq:v}, for both components of the  cytoplasmic flow field in terms of the boundary forcing.

	\subsection{Streamfunction}
	
	For incompressible axisymmetric flows, it is often useful to define the Stokes streamfunction $\psi$ as $u = \tfrac{1}{r}\tfrac{\partial \psi}{\partial r}, v = -\tfrac{1}{r}\frac{\partial \psi}{\partial z}$. This formulation ensures that incompressibility is automatically satisfied, and simplifies problems by reducing the governing equations into a single  equation for $\psi$. In our case, since we have already determined $u$ and $v$, we may determine $\psi$ by integrating the definition of the streamfunction. We thus derive an analytical expression for the streamfunction,
	\begin{equation}\label{eq:psi}
		\psi(r,z,t) = \int_0^{r}r' u(r',z,t)\mathrm dr' = U(z,t)\left(\frac{r^4}{2R(z)^2} - \frac{r^2}{2}\right).
	\end{equation}
	Noting that streamlines are level curves of $\psi$, this allows us to easily plot streamlines of the cytoplasmic flow.
	
	\subsection{Cortical stress}
	Using our cytoplasmic flow solution, we may now calculate the profile of shear stress exerted by the boundary onto the bulk fluid. In the context of cortex-driven cytoplasmic flows~\cite{shamipour2019, fuentes2010,deneke2019, lopez2023,gubieda2020}, this has the direct interpretation as the stress exerted by the cortex onto the cytoplasm, and is a biophysically important quantity in that it informs the active force generation mechanisms in the cell cortex.

	The hydrodynamic stresses due to a flow field are encoded in the Cauchy stress tensor $\boldsymbol \sigma := -p \mathbf I + \mu [\nabla \mathbf u + (\nabla \mathbf u)^T]$. Defining the normal unit vector to the cortex, pointing away  from the cytoplasm, as $\mathbf n = (\mathbf e_r - R'\mathbf e_z)/\sqrt{1 + R'^2}$, and tangent unit vector as $\mathbf t = (R'\mathbf e_r + \mathbf e_z)/\sqrt{1 + R'^2}$, the shear stress $\sigma$ exerted by the cortex onto the cytoplasm is given by $\sigma := \mathbf n \cdot \boldsymbol \sigma \cdot \mathbf t \vert_{r = R}$, which may be expanded as
	\begin{equation}\label{eq:stress1}
		\sigma = \frac{2\mu R'( \partial_rv - \partial_zu) + \mu(1 - R'^2)(\partial_zv + \partial_ru)}{1 + R'^2}\bigg\vert_{r = R}.
	\end{equation}
	We further use our lubrication solution for $u$ and $v$ to provide explicit expressions for the velocity gradients in the numerator, leading to
	\begin{subequations}\label{eq:stress2}
		\begin{align}
			\partial_z u\vert_{r = R} &= U' - \frac{4UR'}{R},\\
			\partial_r u\vert_{r = R} &= \frac{4U}{R},\\ \label{eq:stress2_3}
			\partial_z v\vert_{r = R} &= 2U'R' + UR'' - \frac{3UR'^2}{R},\\ 
			\partial_r v\vert_{r = R} &= -U' + \frac{3UR'}{R}.
		\end{align}
	\end{subequations}
	These equations \eqref{eq:stress1} and \eqref{eq:stress2} specify the shear stress onto the cortex. 
	
	{Note that,} although lubrication theory is a leading order approximation of an asymptotic expansion in the inverse aspect ratio, we have chosen here to keep the denominator unexpanded in equation \eqref{eq:stress1}. This is because the $R'^2$ term is necessary to prevent unphysically large stresses near the poles of the cell, where $R'$ is large. In a strict mathematical sense, we are therefore outside the formal regime of validity of lubrication theory; we will show below that this allows our solution to agree quantitatively with full numerics. 
	
	\subsection{Summary of analytical model}
	This concludes our model for cytoplasmic flows in elongated cells.  {Given a cell geometry $R(z)$ and a boundary velocity $U(z,t)$, we have now determined analytically the flow field inside the cell as well as the cortical stress.} Our main results are fully analytical expressions for the cytoplasmic flow field (equations \eqref{eq:u} and \eqref{eq:v}), the streamfunction (equation \eqref{eq:psi}), and   the cortical stress (equations \eqref{eq:stress1} and \eqref{eq:stress2}). {In what follows, we apply these results to cytoplasmic streaming in different biological cells, showing that classical fluid mechanics provides an accurate description of a complex biological phenomenon.}
	
	\section{Application to \textit{Drosophila} embryos}\label{sec:drosophila}

	\subsection{Motivation}

\begin{figure}[t!]
		\includegraphics[width=\textwidth]{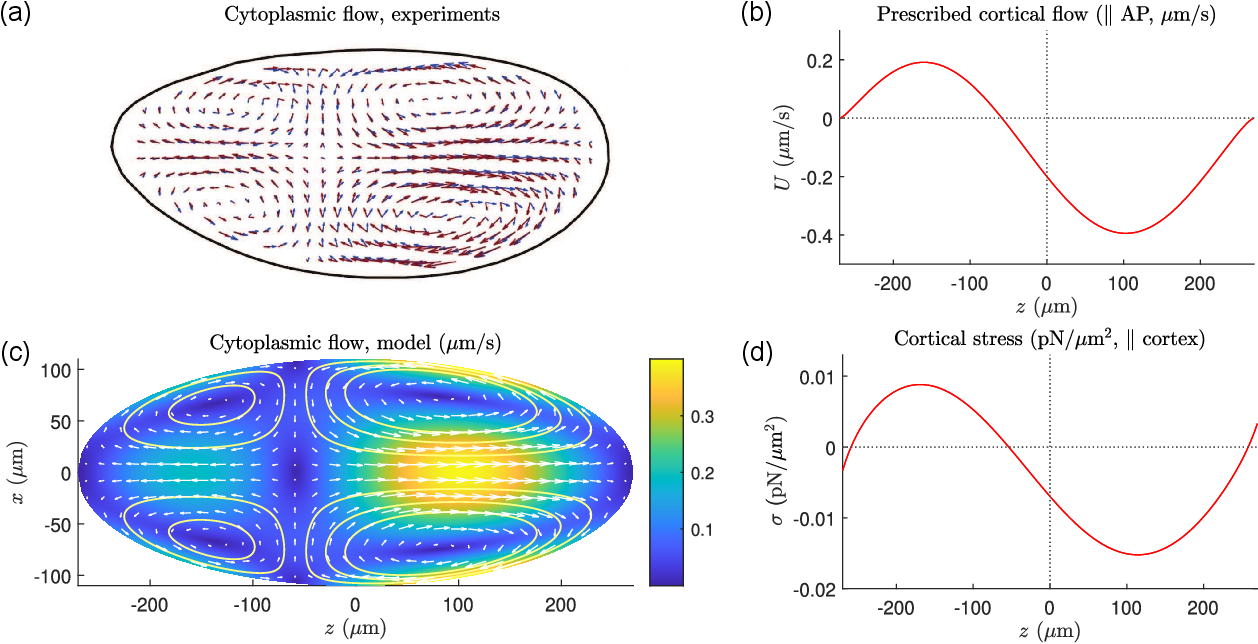}
		\caption{ {Flows and stresses in the \textit{Drosophila} embryo.} (a) Experimentally measured cytoplasmic flow in a \textit{Drosophila} embryo, reproduced from Deneke \textit{et al.}~(2019), \textit{Cell}, {\bf 177}, 925-941~\cite{deneke2019}  {with permission from Elsevier}. (b) Prescribed cortical flow $U(z,t)$, chosen to closely match experimentally measured cortical flows, plotted against longitudinal coordinate $z$, at the contraction peak of cell cycle 6. (c) Cytoplasmic flow field reconstructed using lubrication solution. The velocity is plotted as white arrows on a regular grid; the color map indicates speed and pale yellow curves are streamlines. (d) Shear stress $\sigma$ exerted onto cortex, plotted against $z$.}\label{fig:D}
	\end{figure}

	We first apply our methods to cortex-driven cytoplasmic flows in the syncytial \textit{Drosophila} embryo, a large, elongated cell which contains numerous nuclei in a common cytoplasm. After the oocyte is fertilized, the nucleus undergoes 14 rounds of cell division, termed ``cell cycles". It has been shown experimentally that cytoplasmic flows during cell cycles 4 to 6 driven by cortical contractions play a crucial role in spreading the nuclei throughout the embryo, and that a uniform nuclear distribution is required for proper embryonic development~\cite{deneke2019}. {The cortical flows are bidirectional and directed from a slightly off-center $z$-position towards the anterior and posterior poles (on the left and right respectively oriented as shown in Fig.~\ref{fig:D}). The experimentally measured cytoplasmic flows, consisting of four vortices (as visualized in a cross-section through the anterior-posterior (AP) axis)} which push nuclei near the center along the AP axis towards the poles, are reprinted from Ref.~\cite{deneke2019} in Fig.~\ref{fig:D}a. We have modeled these flows in a previous work~\cite{htet2023} and demonstrated there that the real-life cortical flows enable near-optimal spreading of nuclei. We briefly summarize here the results on the cytoplasmic flow field as calculated using lubrication theory, as a preliminary demonstration of the impact of the analytical modeling approach, and further proceed to determine the cortical stress.

	\subsection{Cytoplasmic flows and cortical stresses}
	
	The \textit{Drosophila} embryo is approximately a prolate spheroid of semi-major axis $L = 270$~\textmu{m} and semi-minor axis $R_0 = 110$~\textmu{m}. Using the notation in \S\ref{sec:maths}, the cell geometry is thus described by 
	\begin{equation}
		R(z) = R_0\sqrt{1 - \left(\frac{z}{L}\right)^2}.
	\end{equation}
	The cortical flows in cell cycles 4 to 6 have a complex spatial and temporal dependence, but to a good approximation, may be modeled as the product of a time-varying amplitude $V(t)$ and a sinusoidal spatial profile, $\mathcal U(z,t) = V(t){\left[-\tfrac{1}{3} - \sin\left(\tfrac{\pi z}{L} + \arcsin\tfrac{1}{3}\right)\right]}$~\cite{htet2023}. The $z$-component $U(z,t) = \mathcal U(z,t)/\sqrt{1 + R'(z)^2}$ is plotted in Fig.~\ref{fig:D}b at the contraction peak of cell cycle 6, at which instant $V(t)$ takes a numerical value of $0.3$~\textmu{m}/s. 
	
	Now that we have specified $U(z,t)$ and $R(z)$, the flow field is obtained from equations \eqref{eq:u} and \eqref{eq:v}, and plotted in Fig.~\ref{fig:D}c; white arrows indicating the magnitude and direction of the velocity field are superposed onto a color map of the speed. Streamlines are plotted in pale yellow in the same figure using our expression for the streamfunction (equation \eqref{eq:psi}). These results are in good agreement with experimental measurements (Fig.~\ref{fig:D}a).
	
	Importantly, these lubrication results have been validated in Ref.~\cite{htet2023} against an exact solution using spheroidal harmonics, and shown to incur a maximum relative error of 5\%, defined for each velocity component $u$ and $v$ as the maximum absolute difference between the lubrication and exact solution relative to the maximum value attained by the exact solution. This validation illustrates the  excellent accuracy of {our lubrication model compared to exact Stokes flow solutions}, and gives us the confidence to further apply this to other biological systems.

	We may further use the theoretical model to infer the shear stress exerted by the cortex (equations \eqref{eq:stress1} and \eqref{eq:stress2}), and plot this stress profile in Fig.~\ref{fig:D}d. The stress is in the same direction as the cortical flows and near-cortex cytoplasmic flows, consistent with the physical picture of active cortical stresses driving flow. Our model therefore provides a simple tool to probe the cortical stress, allowing us to circumvent the experimental challenges of direct cortical stress measurements and infer the cortical stress from only cortical flow measurements.  {These stresses could, in turn, be used to infer biochemical information, such as myosin gradients.}

		{The primary biological function of the cytoplasmic flows is to spread the numerous nuclei inside the \textit{Drosophila} embryo along the embryo's long axis, ensuring the homogeneous nuclear distribution necessary for correct embryonic development. In Ref.~\cite{htet2023}, the flow field solutions were further employed in transport simulations to  show that the biological flows are finely tuned to achieve a near-optimal amount of nuclear spreading. If the cytoplasmic flows are too weak, the nuclei do not spread adequately away from their initial positions near the embryo's center, and if the flows are too strong, an inhomogeneous nuclear distribution results, with nuclei accumulating at the poles. In Ref.~\cite{lopez2023}, numerical solution of a more complicated model of the flow was used to obtain important biological insight on the gradient of Bicoid, a morphogen responsible for organization of anterior development in \textit{Drosophila}~\cite{driever1988}. By solving transport equations with and without flow, the authors refuted the hypothesis~\cite{hecht2009} that flow might play a role in establishing this gradient, and the analytical solutions we have developed here may also be used in a similar manner to investigate biological hypotheses. These examples emphasise that the value of our model lies not only in its ability to reproduce measured or computed flow fields and stresses, but also, equally importantly, in the potential of these results and solutions to be directly used by the biological community to test specific hypotheses.}

	\section{Application to \textit{C.~elegans} embryos}\label{sec:celegans}
	
	\subsection{Motivation}
	In this section, we apply our model to cytoplasmic flows in the \textit{C.~elegans} embryo at the one-cell stage. In this system, the cortical flows are roughly unidirectional and directed towards the anterior pole, and this drives a cytoplasmic flow which is directed towards the posterior pole along the AP axis and towards the anterior pole near the cortex~\cite{niwayama2011}. These flows help distribute cytoplasmic and cortical cellular components asymmetrically and are thought to be important for establishing cell polarity~\cite{gubieda2020}. {In order to understand  {actomyosin dynamics} in the cortex and the biophysical origins of these flows, it is important to quantify the cortical stresses. 
		
		\begin{figure}[t!]
			\includegraphics[width=0.6\textwidth]{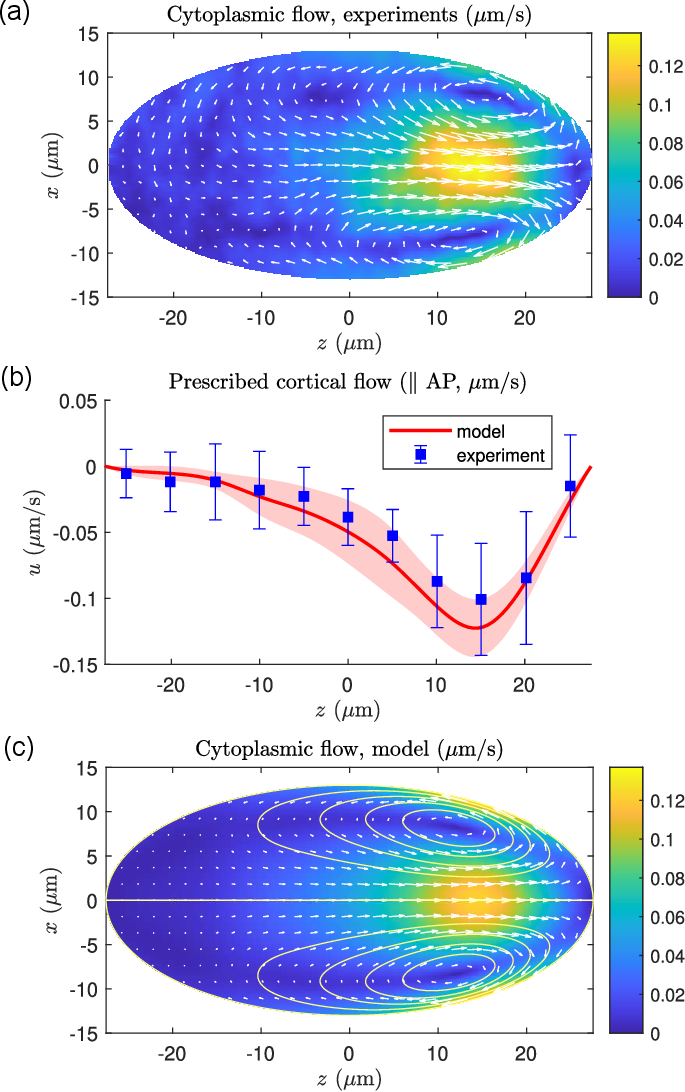}
			\caption{{Cytoplasmic and cortical flows in the \textit{C.~elegans} embryo. }(a) PIV data of cytoplasmic flow field in the \textit{C.~elegans} embryo,  {measured in Niwayama \textit{et al.}~(2016), \textit{PLOS ONE}, {\bf 11}, e0159917~\cite{niwayama2016}}, averaged over six embryos. (b) Prescribed cortical boundary condition (red), fitted to minimize least square errors with PIV data in the six embryos, compared with experimentally measured cortical flows (blue). Shaded area indicates mean $\pm$ standard deviation across the six embryos. (c) Cytoplasmic flows corresponding to this fitted cortical flow. Color indicates flow speed and streamlines are in white; the centerline flow is in the positive $z$ direction.}\label{fig:Cflows}
		\end{figure}
		
		In a previous work, cortical stresses were inferred by fitting cytoplasmic flow solutions from full hydrodynamic simulations to experimentally measured values using Bayesian data assimilation~\cite{niwayama2016}.  In what follows, we demonstrate that these flows and stresses can be reproduced by our analytical model and a least-squares fit, without the need for sophisticated computational methods. We then illustrate the robustness of our model to cell shapes by characterizing flows when the embryo exhibits a partial constriction called a pseudocleavage furrow.}

	\subsection{Cell geometry and experimental cytoplasmic flows}
	
	In Ref.~\cite{niwayama2016}, cytoplasmic flows were measured in six embryos. Although the embryos are approximately spheroidal, their precise shapes are irregular and exhibit natural variation across different embryos. The hydrodynamic simulations were performed in a spherocylindrical domain {(i.e.~a cylinder with hemispherical caps)} whose long axis is 55/13 ($\approx 4.23$) times its short axis, reflecting biological \textit{C.~elegans} embryo dimensions. The positions of the velocity data points taken in each embryo were linearly rescaled into the same spherocylindrical shape to facilitate the fitting procedure and standardize the shape variation across embryos (see Methods in Ref.~\cite{niwayama2016} for details).

	Here, we perform a similar rescaling of the six velocity fields reported in Ref.~\cite{niwayama2016} into a prolate spheroid of semi-major axis $L = 27.5$~\textmu{m} and semi-minor axis $R_0 = 13$~\textmu{m}. A spherocylinder has a discontinuity in curvature at the sphere-cylinder transition, and thus introduces an unphysical jump in stress (via a discontinuity in the  $R''$ term in equation \eqref{eq:stress2_3}); we have chosen the smooth geometry of a prolate spheroid in order to prevent {this artificial mathematical singularity. } {We note, however, that this discontinuity is integrable and thus the total force on any section of the cortex remains finite even when a spherocylinder is used.}

We thus choose to work in a prolate spheroidal cell, and first visualize the experimentally measured flows, averaged across the six embryos, in Fig.~\ref{fig:Cflows}a; here, white arrows indicate direction and magnitude of the velocity field, and the background color map indicates iso-values of the speed. The grid points at which the velocity data in Ref.~\cite{niwayama2016} is taken are irregular and vary across embryos; in order to take an average across the embryos, we have bilinearly interpolated the experimental data onto a regular Cartesian grid. The resulting flow field, as visualized in this 2D cross-section through the AP axis, has two vortices centered in the posterior (right) half of the embryo.

\subsection{Fitting  {procedure for} cortical flows}

Our aim is to determine the cortical flow profile $U(z)$ which, using our  model, produces cytoplasmic flows that best match experimental measurements. We first constrain the functional form of $U(z)$ to be a spline interpolation through the eleven points $(z_i, U_i)$ for $i = 1, \dots, 11$, with $z_i = -L + \tfrac{2L}{11}i - L/11$ fixed, and the $U_i$'s to be determined. In Ref.~\cite{niwayama2016}, the stress exerted by the cortex is constrained on physical grounds to be unidirectional, pointing from the posterior to the anterior, and to be zero at the poles. Using MATLAB's nonlinear constrained minimization routine \texttt{fmincon.m}, we   determine for each embryo the values of $U_i$ which minimize in a least squares sense, subject to the same cortical stress constraint, the error between the experimental measurements and the lubrication cytoplasmic flow solution. 

\subsection{{Prediction of} cytoplasmic flows and stresses} 

\begin{figure}[t!]
	\includegraphics[width=\textwidth]{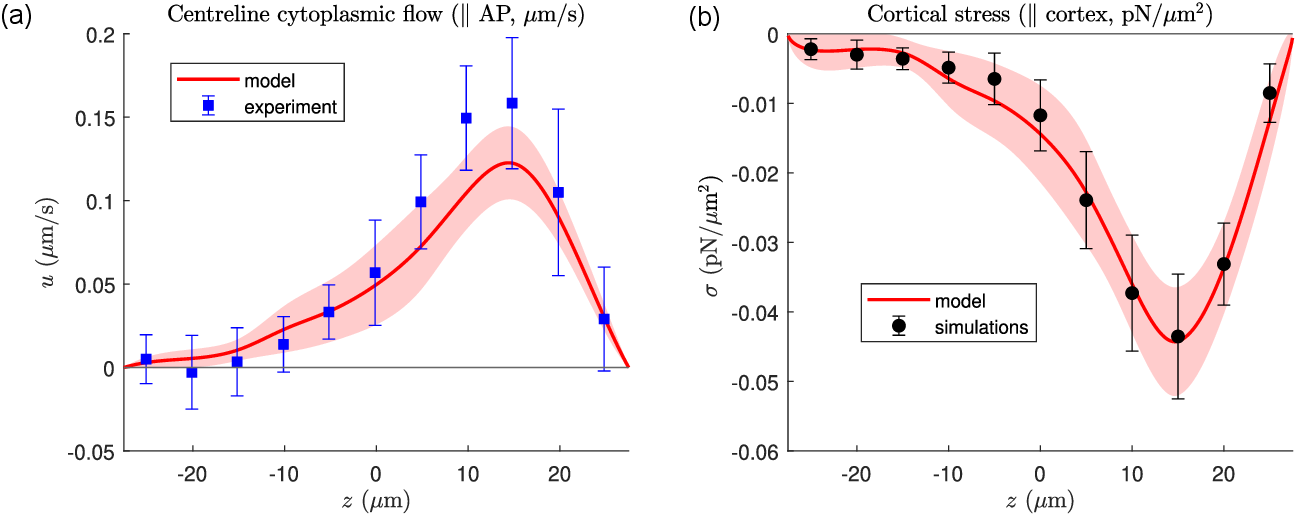}
	\caption{{Model predictions for centerline flow and cortical stress in the \textit{C.~elegans} embryo.} (a) Centerline cytoplasmic flow in the \textit{C.~elegans} embryo as calculated from our model (red) and as measured experimentally (blue) in  Niwayama \textit{et al.}~(2016), \textit{PLOS ONE}, {\bf 11}, e0159917~\cite{niwayama2016}. (b) Tangential cortical stress as calculated from model (red) and as computed using  simulations and Bayesian inference techniques (black) in Ref.~\cite{niwayama2016}.  In both panels, shaded areas indicate mean $\pm$ standard deviation across the six embryos.}\label{fig:Cstress}
\end{figure}

We plot in Fig.~\ref{fig:Cflows}b the average over the six embryos of the fitted cortical velocities $U(z)$ (red solid line), with the shaded area indicating $\pm$ one standard deviation. We also show in blue the cortical velocity measured experimentally in Ref.~\cite{niwayama2016}, and see {quantitative} agreement with our results. The cytoplasmic flow field corresponding to our fitted cortical flow (Fig.~\ref{fig:Cflows}c) matches experiments (Fig.~\ref{fig:Cflows}a) similarly well.  Next, we plot the cytoplasmic velocity at the cell's centerline (Fig.~\ref{fig:Cstress}a, red) as predicted by our model, against experimental measurements in blue; here also, we see that the analytical model is able to quantitatively capture the spatial dependence of the centerline flow. {Although we would not expect a precise match between the data and our model due to challenges in imaging flows at this scale and inherent biological noise and complexity, our lubrication model is able to capture the important features of the flow.}

The main result of this section concerns the inference of cortical stresses. 
We  compute the  stress profile as predicted from our analytical   model and plot the result  in  	Fig.~\ref{fig:Cstress}b; the  red solid line shows the mean over the six embryos and the shaded area the standard deviation. These predictions are superimposed on the numerical simulations from Ref.~\cite{niwayama2016}, obtained using a combination of full Stokes computations and Bayesian inference techniques (black dots and error bars). 
{We see that our model agrees quantitatively} with the cortical stress calculated using full numerical approaches. Using our analytical solution and MATLAB's minimization routine, the fitting procedure used to produce our results typically takes less than 10 to 20 seconds for each embryo on a standard laptop computer. 	
{Our model thus provides  an effective and inexpensive method to infer cortical stress from flow data alone}, bypassing computationally intensive simulations and the myriad challenges associated with direct experimental stress measurements.

{Beyond an estimation of cortical stresses, an important topic that our flow solution could be used to investigate is the role of flow in the localization of membrane-less organelles called germ granules (referred to as P granules in {\it C.~elegans}) in the posterior half of the embryo~\cite{brangwynne2009}. Experimental work  showed that this localization is caused by dissolution of granule components in the anterior half and condensation  in the posterior region, rather than flow-based migration~\cite{brangwynne2009}. However, a more detailed investigation on the interplay between flow and phase change could reveal a more nuanced understanding of the role of flow.
}

\subsection{Flows in an embryo during pseudocleavage}	

\begin{figure}[t!]
	\includegraphics[width=0.6\textwidth]{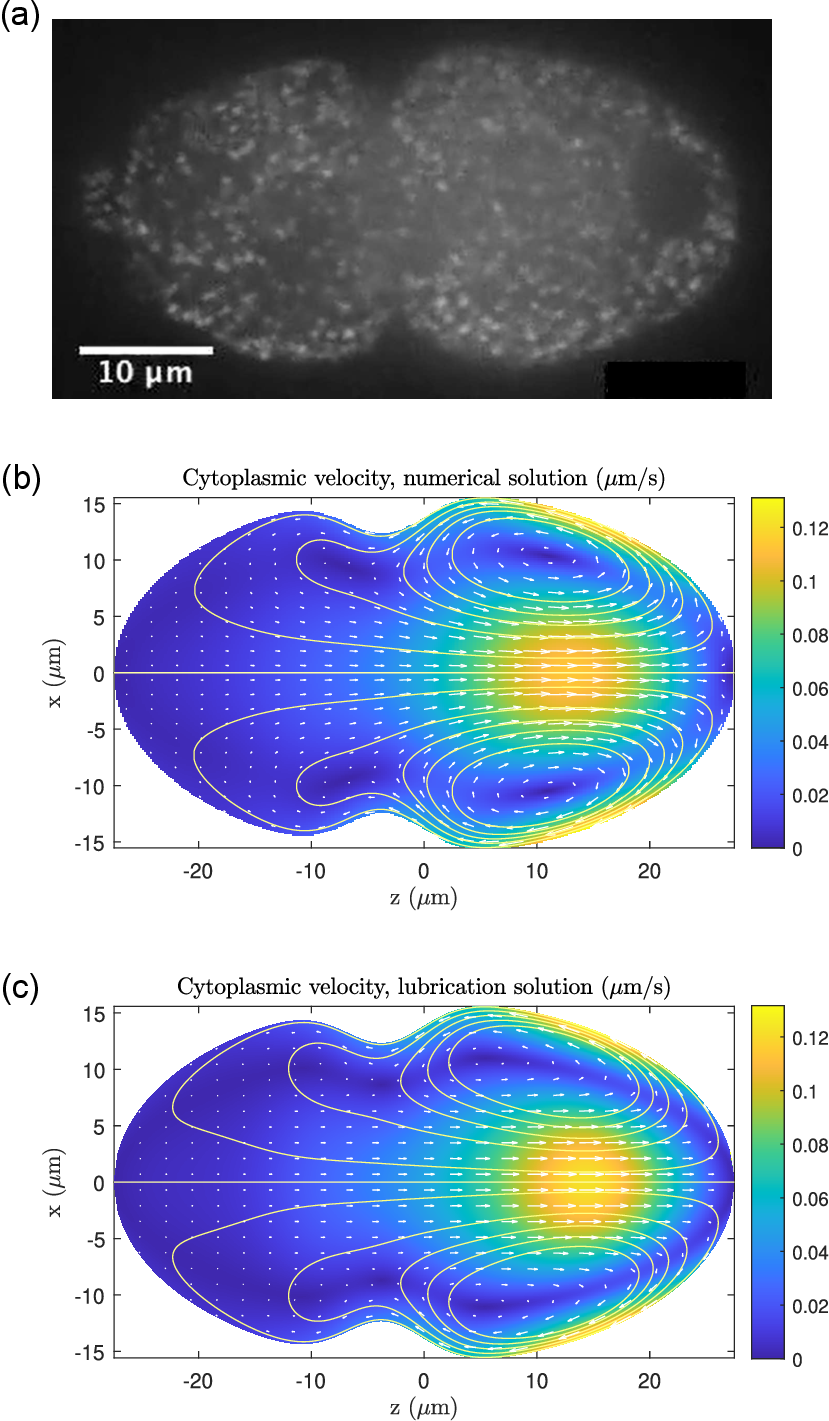}
	\caption{{Cytoplasmic flows in the \textit{C.~elegans} embryo during pseudocleavage.} (a) Microscopy image of a \textit{C.~elegans} embryo during pseudocleavage, reproduced from Supplementary Video 1 in  Niwayama \textit{et al.}~(2016), \textit{PLOS ONE}, {\bf 11}, e0159917~\cite{niwayama2016}. 
		(b,c) Cytoplasmic flow field {and streamlines} in embryo 
		as computed numerically using COMSOL Multiphysics\textregistered~5.6 (b) and our lubrication model (c) with  geometry  fitted to that in (a).}\label{fig:Ccleavage}
\end{figure}

In this section, we further  illustrate a particular strength of our lubrication model: its applicability to domains with more complex geometries. 
As part of embryonic development, the \textit{C.~elegans} embryo undergoes a process called pseudocleavage, in which the cortex undergoes contractions and forms a partial constriction of the embryo called a pseudocleavage furrow~\cite{cuenca2003}. In Fig.~\ref{fig:Ccleavage}a  we reproduce a microscopy image of a \textit{C.~elegans} embryo during pseudocleavage from Supplementary Video 1 of Ref.~\cite{niwayama2016}. 
Pseudocleavage has been found to be largely dispensable for normal embryonic development~\cite{rose1995},  {and its biological significance is not fully understood, but for instance, it has been shown to be important for convergence of the AP axis towards the geometrical axis of the surrounding eggshell when these axes are misaligned~\cite{bhatnagar2023}.} Cortical and cytoplasmic flows continue to occur in the presence of a pseudocleavage furrow, and we now show how our model can be used to characterize flows in this geometry.

We model the embryo shape during pseudocleavage as a spheroid of semi-major axis $L$ and semi-minor axis $R_0$, with a Gaussian-shaped   furrow of depth $d$ and characteristic width $\ell$, centered at a longitudinal position $z_0$
\begin{equation}
	R(z) = \left\{R_0 - d\exp\left[-\frac{(z-z_0)^2}{2\ell^2}\right]\right\}\sqrt{1 - \frac{z^2}{L^2}}.
\end{equation}
The parameters $R_0, L, z_0, d,$ and $\ell,$ are chosen as the best fit to the experimental image in Fig.~\ref{fig:Ccleavage}a. Specifically, we threshold the experimental image into a binary image whose pixels take a value of 0 inside the embryo and 1 outside, generate a similar binary image using our chosen form of $R(z)$, and determine the parameter values which minimize the absolute difference between these two images; the resulting fitted shape is seen in 
Figs.~\ref{fig:Ccleavage}b-c.

In the absence of detailed flow measurements during pseudocleavage, we prescribe the cortical velocity $U(z)$ determined in the previous subsection (and illustrated in Fig.~\ref{fig:Cflows}b). The corresponding cytoplasmic flows {(and streamlines)} as computed using COMSOL Multiphysics\textregistered~5.6 and our lubrication model are plotted in Fig.~\ref{fig:Ccleavage}b and  Fig.~\ref{fig:Ccleavage}c respectively. Despite the presence of the furrow, the analytical model is in {very good}  agreement with the full numerical simulations. 
We may quantify this by calculating a RMS error of 18\%, defined as $\left[\int |\mathbf u_{\text{sim}} - \mathbf u_{\text{model}}|^2\,\mathrm dA\right]^{1/2}/\left[\int |\mathbf u_{\text{sim}}|^2\,\mathrm dA\right]^{1/2}$, where the integrals are evaluated over the entire 2D cross-section illustrated in Figs.~\ref{fig:Ccleavage}b-c. {Lubrication is known to be accurate to quadratic error in the slenderness, defined loosely for our cell as the ratio of the radial length scale to the longitudinal length scale. Since the relevant radial length scale is \textit{half} the width of the cell (keeping in mind the axisymmetry) and the longitudinal length scale is, due to the pseudocleavage, around half the cell length, the slenderness ratio for this system is approximately $0.5$. This corresponds to an error on the order of $0.5^2 = 25$\%, an order-of-magnitude estimate that is consistent with the 18\% RMS error  calculated above.} Despite the pseudocleavage introducing a new longitudinal length scale which is smaller than the embryo length, lubrication theory is able to capture all relevant physical features of the flow.

\clearpage

\section{Solution in an annular domain and application to pollen tubes}\label{sec:pollentubes}

\subsection{Motivation}

In this section, we show that our model   can be applied to the pollen tubes. A pollen tube is a long protrusion from a pollen grain which transfers sperm cells to the ovary of the receiving plant, and is another example of a cell in which cytoplasmic streaming performs important transport functions~\cite{chebli2013}. It is a long cylindrical cell with an approximately hemispherical cap. {In angiosperms (flowering plants)} vesicles and other cargo are transported actively towards the tip by molecular motors along actin bundles running along the peripheral wall, and away from the tip along a central actin bundle~\cite{qu2017, xu2020, zhang2023}.  {The consequent fluid flows are known as `reverse fountain streaming', and contribute further transport via passive advection of cellular constituents}. {Note that the pollen tubes of gymnosperms also exhibit fluid flows; they are in the opposite direction to the flows in angiosperms and known as `fountain streaming'~\cite{chebli2013}. In the following subsections, we will focus on the pollen tubes of the lily, an angiosperm.}

.  

\begin{figure}[t!]
	\includegraphics[width=0.95\textwidth]{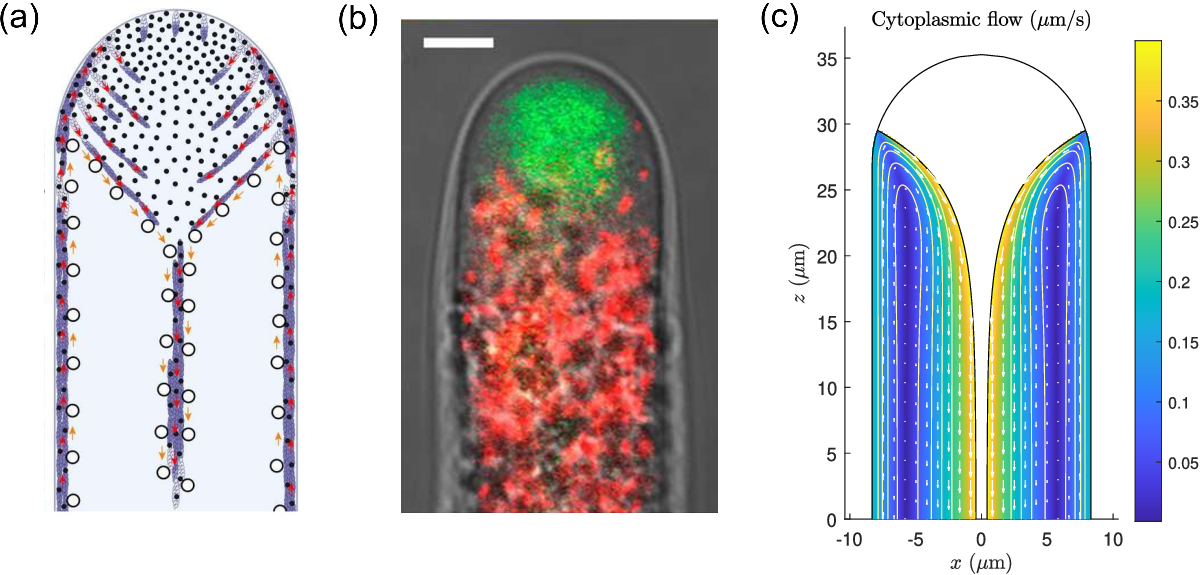}
	\caption{{Cytoplasmic flows in the pollen tube.} (a) Schematic showing organization of peripheral and central actin bundles in a pollen tube, and active transport of vesicles along the actin bundles, reproduced from {Zhang \textit{et al.}~(2023), \textit{Plant Physiol.}, \textbf{193}, 9~\cite{zhang2023}} {with permission from Oxford University Press}. (b) Fluorescence micrograph of a lily pollen tube, with vesicles stained in red, reproduced from {Bove \textit{et al.}~(2008), \textit{Plant Physiol.}, \textbf{147},
		1646~\cite{bove2008}.} Scale bar is 5~\textmu{m}. (c) Cytoplasmic flow field predicted by our model. White arrows indicating velocity vector field are overlaid onto a color map of flow speed.}\label{fig:P}
\end{figure}

\subsection{Cell geometry}

A schematic representation of actin filament organization and vesicle transport along actin bundles in a pollen tube is shown in Fig.~\ref{fig:P}a.  {Actin filaments are present much more densely in the apical region, which prevents the entry of larger organelles such as mitochondria~\cite{zhang2023}.}   Consistent with this picture, we reproduce in Fig.~\ref{fig:P}b   a fluorescence micrograph of a lily pollen tube in which vesicles are stained in green and mitochondria in red~\cite{bove2008}, and we see a clear division between the apical region and the shank region.

Motivated by this setup, we consider the cytoplasmic flow in an annular domain of inner radius $R_0(z)$ and outer radius $R_1(z)$.  We model the entrainment of cytoplasm due to active transport of molecular cargoes along the actin bundles as prescribed tangential slip velocities along both walls. We denote the $z$-components of the slip velocities at the inner and outer walls by $U_0(z,t)$ and $U_1(z,t)$ respectively. 
{We derive in the next section a cytoplasmic flow solution applicable to general annular geometries and slip velocities. Subsequently, we focus on the pollen tube geometry by taking $R_1(z)$ to be a spherocylinder to specify the overall shape of the pollen tube and using an exponential curve for $R_0(z)$ to model the separation between the cytoplasm in the shank region and the actin-dense apical region/central actin bundle.}

\subsection{Cytoplasmic flow  {solution in an annular geometry}}

Assuming that the growth rate of the tube is negligible relative to the typical vesicle speed, and that cytoplasmic flows into or out of the dense tip region are sufficiently small  that there is zero volume flux of cytoplasm in the solution domain, it is straightforward to  solve the longitudinal component of Stokes' equations in the annular domain, subject to the prescribed velocity boundary conditions at the inner and outer walls, to determine the longitudinal component $u(r,z)$ of the flow field,
\begin{equation}\label{eq:p1}
	u(r,z,t) = G(z,t)[r^2 - R_1(z)^2] + A(z)\ln\frac{r}{R_1(z)} + U_1(z,t),
\end{equation}
where $A(z,t)$ and $G(z,t)$ satisfy
\begin{align}\label{eq:p2}
	\begin{pmatrix}
		R_0^2 - R_1^2 & \ln\left(\frac{R_0}{R_1}\right) \\
		-\frac{1}{4}(R_0^2 - R_1^2)^2 & \frac{1}{4}R_0^2\left[1 - 2\ln\left(\frac{R_0}{R_1}\right)\right]-\frac{1}{4}R_1^2
	\end{pmatrix}\begin{pmatrix}
		G \\ A
	\end{pmatrix} = \begin{pmatrix}
		U_0 - U_1 \\ \frac{1}{2}U_1(R_0^2 - R_1^2)
	\end{pmatrix}.
\end{align}
Should one wish to account in more detail for the growth of a pollen tube, the velocity boundary conditions may be appropriately modified and the no-flux condition relaxed, although we do not consider this level of detail in the current study. 

Similarly to before, the radial component of the velocity, $v(r,z,t)$,  may next be calculated by integrating the incompressibility condition $\nabla \cdot \mathbf u = 0$. This time, the integration constant is determined by requiring $v$ to satisfy the slip velocity condition at one of the walls.  {Since the prescribed slip is tangential to the cell surface which itself is, in general, locally not parallel to the longitudinal axis, there is a radial component of the slip boundary condition. For ease, we apply this radial boundary condition $v(r = R_1) = U_1(z,t)R_1'(z)$ at the outer wall, and the slip velocity condition at the inner wall is then necessarily satisfied by integrating the incompressibility condition. We therefore obtain}  	\begin{align}\label{eq:p3}
	v(r,z) &= -G'(z,t)\left[\frac{r^3}{4} - \frac{rR_1(z)^2}{2} + \frac{R_1(z)^4}{4r}\right] - \frac{A'(z,t)}{4}\left[r\left(2\ln\frac{r}{R_1(z)} - 1\right) + \frac{R_1(z)^2}{r}\right] \\ \nonumber
	&+\left[G(z,t)R_1(z)R_1'(z) + \frac{A(z,t)R_1'(z)}{2R_1(z)} - \frac{U_1'(z,t)}{2}\right]\left[r - \frac{R_1(z)^2}{r}\right] + \frac{U_1(z,t)R_1'(z)R_1(z)}{r}.
\end{align}
We may further show that the streamfunction is 
\begin{equation}
	\psi = \frac{G(z,t)}{4}\left[R_1(z)^2 - r^2\right]^2 + \frac{A(z,t)}{4}\left[2r^2\ln\left(\frac{r}{R_1(z)}\right) - r^2 + R_1(z)^2 \right] + \frac{U_1(z,t)}{2}\left[r^2 - R_1(z)^2\right].
\end{equation}

\subsection{Cytoplasmic flow prediction}
To the best of our knowledge, no complete measurements of the cytoplasmic flow field in a pollen tube are available in the literature, although vesicle transport has been measured in varying levels of detail~\cite{bove2008, hu2017}. Using the vector map of organelle movement reported in Ref.~\cite{bove2008}, we take the approximate transport speeds along the peripheral and central actin bundles as 
0.2~\textmu{m}/s and 0.4~\textmu{m}/s respectively --  {specifically, we set the tangential velocities along the inner and outer boundaries to these constant values}. The same work reports a typical value of 8.3~\textmu{m} for the radius of the pollen tube.
We illustrate our model geometry, and the predicted cytoplasmic flows  (equations \eqref{eq:p1},\eqref{eq:p2}, and \eqref{eq:p3}), in Fig.~\ref{fig:P}c. {These flows may be used in transport equations to study transport phenomena in pollen tubes, {for instance, of calcium ions}. The simplicity of the final results illustrates the predictive power of this modeling approach and may be refined upon availability of more detailed experimental measurements.
	
	\section{Azimuthal flows and application to root hair cells}\label{sec:roothaircells}
	
	\subsection{Motivation}
	
	Thus far, we have considered cytoplasmic flows with no chiral component and seen how our results can be used to model several distinct cells. In this section, we further generalize our model to incorporate azimuthal flows, thus allowing us to characterize a wider range of biological systems.

	{Root hairs are long cylindrical extensions from root epidermal cells, which greatly increase the surface area of the plant root system and thus facilitate nutrient acquisition, anchorage and microbe interactions~\cite{grierson2014}.}
	Similar to pollen tubes, plant root hair cells contain thin actin bundles at the cell periphery and thick transvacuolar actin bundles at the centerline which transport vesicles and other organelles~\cite{tominaga2000}, and exhibit reverse fountain streaming in the cytoplasm-dense region towards the tip~\cite{sieberer2000}. The plant hormone auxin plays important roles in plant growth and development~\cite{gomes2001}, including the regulation of cytoplasmic streaming, and has been reported to enhance cytoplasmic streaming at low concentrations and inhibit it at high concentrations. In order to understand the mechanisms behind this inhibition of cytoplasmic streaming, Tominaga \textit{et al.}~\cite{tominaga1998} studied root hair cells of the aquatic plant \textit{Hydrocharis}, subjected to high concentrations of the synthetic auxin, naphthalene acetic acid (NAA). This acidification disrupts the organization of actin filaments, thus inhibiting cytoplasmic streaming, but the cell recovers to a normal reverse fountain streaming state several hours after removal of NAA. Intriguingly, the authors report a state of helical cytoplasmic streaming during recovery, in which the peripheral actin bundles are helically (rather than longitudinally) oriented and the central bundles have not yet recovered; these empirical observations are  illustrated in Fig.~\ref{fig:R}a. 
	
	In what follows we will illustrate how our analytical model can also be used to characterize this NAA-induced helical streaming. Although the experimental study concerns the inhibition of cytoplasmic streaming \textit{in vitro} via NAA treatments, acidification of the cytoplasmic via auxin influxes also occurs physiologically~\cite{tominaga1998} and we expect our results to be more generally relevant beyond this specific \textit{in vitro} setting. Helically oriented actin arrays are also naturally found in healthy root hair cells of some species such as \textit{Arabidopsis}~\cite{sheahan2004}, and corresponding helical streaming is expected to occur. 
	
	\subsection{Cytoplasmic flow model}
	
	We return to the geometry of an elongated cell with one boundary described by $R(z)$, but first consider  {solely} an axisymmetric azimuthal wall velocity of the form  {$W(z,t)\mathbf e_\theta$, where $\mathbf e_\theta$ is the azimuthal basis unit vector in cylindrical polar coordinates $(r, \theta, z).$} We note that 
	\begin{equation}\label{eq:azi}
		\mathbf u(r,z,t) = \frac{W(z,t)r}{R(z)}\mathbf e_\theta
	\end{equation}
	is the lubrication solution for the resultant flow. This represents solid body rotations of each `slice' of the cell. We work with this cylindrical (rather than annular) geometry, since the central actin bundles in the root hair cells remain disrupted while helical streaming occurs.
	
	Since Stokes flows are linear in the boundary conditions, we may now derive the lubrication solution for a more general driving wall velocity of the form $\mathbf U(z,t) = \mathcal U(z,t)\mathbf t(z) + W(z,t)\mathbf e_\theta$ (where $\mathbf t$ is the unit tangent vector with zero azimuthal component) as the superposition of this purely azimuthal flow, equation \eqref{eq:azi}, with the solution derived in \S\ref{sec:maths} for a tangential slip velocity with no azimuthal component, \eqref{eq:u} and \eqref{eq:v}. In other words, provided the problem is axisymmetric, our model is now able to also characterize chiral cytoplasmic flows.

	\begin{figure}[t!]
		\includegraphics[width=0.95\textwidth]{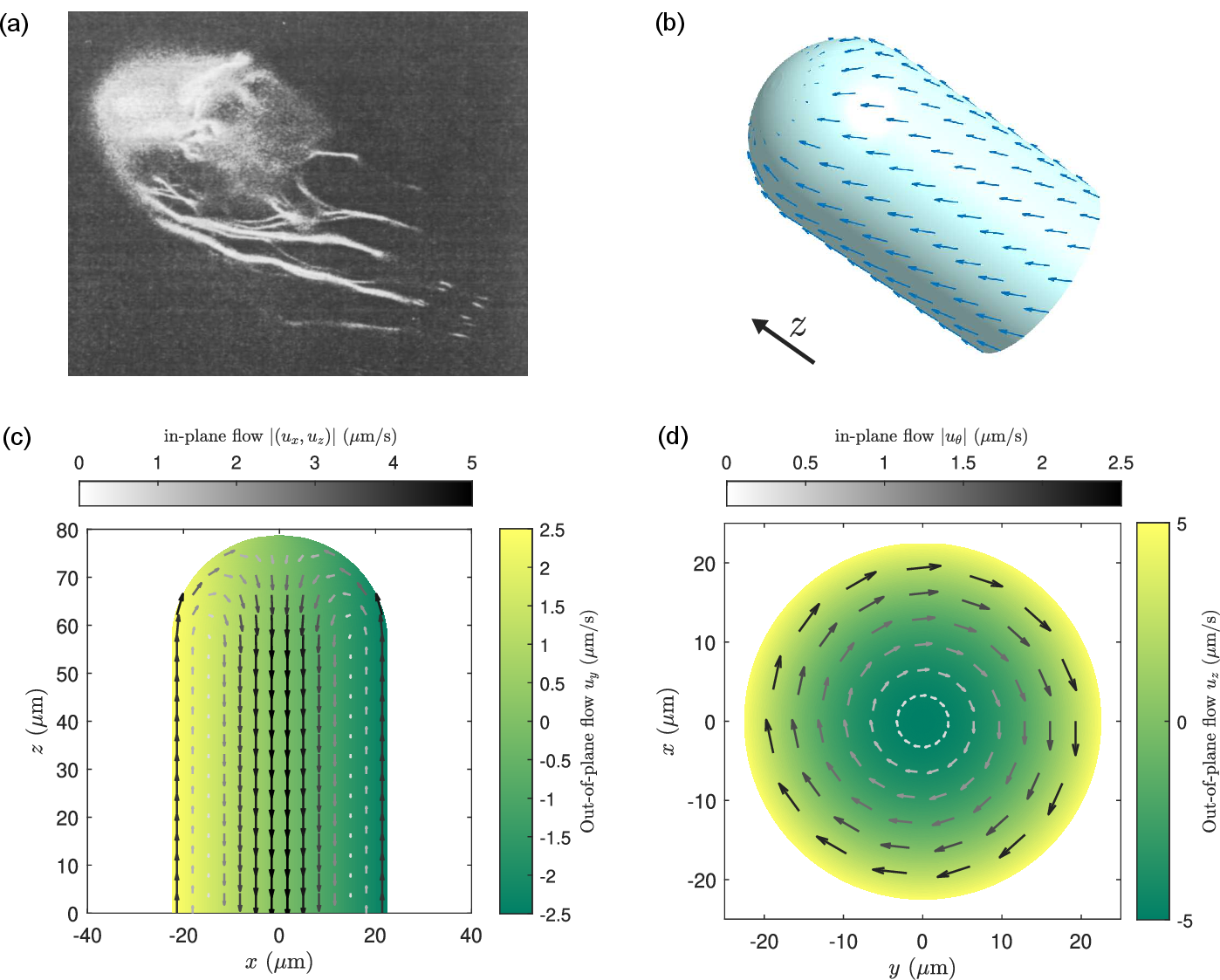}
		\caption{{Cytoplasmic flows in the root hair cell.} (a) Micrograph of actin filaments (white) in a root hair cell undergoing helical streaming, reproduced from  {Tominaga \textit{et al.}~(1998), \textit{Plant  Cell Physiol.}, \textbf{39}, 1342-1349}~\cite{tominaga1998}  {with permission from OUP}. (b) Spherocylindrical model geometry, and prescribed wall velocity illustrated as dark blue arrows. (c, d) Cytoplasmic flow field reconstructed using lubrication solution, illustrated in a cross-section through the longitudinal axis (c) and in a circular cross-section perpendicular to the longitudinal axis viewed from the base (d). In each figure, greyscale arrows indicate in-plane velocity with speeds color-coded according to the horizontal color bar while the background color map indicates out-of-plane component of flow, with positive values (yellow) indicating flow into the plane. }\label{fig:R}
	\end{figure}
	
	\subsection{Cytoplasmic flow predictions}
	We now apply this solution to helical streaming in root hair cells. We model the cell geometry $R(z)$ as a cylinder with a hemispherical cap. Although a spherocylinder has a discontinuity in $R''(z)$, this is not a problem here since we proceed to calculate only the leading order lubrication flow, and in particular, not the boundary stress. Along the cylindrical region, we impose an obliquely oriented slip velocity $\mathcal U \mathbf t  -\frac{1}{2}\mathcal U \mathbf e_\theta$. The magnitude and direction of the azimuthal component are chosen to emulate the experimental helical actin organization illustrated in Fig.~\ref{fig:R}a. Vesicles are released in the tip region via exocytosis, and we therefore expect the flow forcing at the tip to be zero.  {Thus, we set the slip velocity in the apex to be  $\frac{R(z)}{R_{\text{cyl}}}(\mathcal U \mathbf t(z) -\frac{1}{2}\mathcal U \mathbf e_\theta)$, a scaling of the slip velocity in the cylindrical shank region by the ratio $\frac{R(z)}{R_{\text{cyl}}}$ of the local radius $R(z)$ in the apex to the radius $R_{\text{cyl}}$ of the cylindrical region.} The model cell geometry and the imposed wall velocity are illustrated in Fig.~\ref{fig:R}b.  {We further choose a numerical value $\mathcal U = 5$ \textmu{m}/s, a typical value of the helical streaming velocity measured in Ref.~\cite{tominaga1998}.}

	In Fig.~\ref{fig:R}c, we plot the flow field we predicted in a cross-section through the longitudinal axis. The in-plane components (arrows, color-coded according to speed, see horizontal color bar) show reverse fountain streaming (which occurs naturally without helical forcing), and are superposed on a color map of the out-of-plane, i.e.~azimuthal, component. In Fig.~\ref{fig:R}d, we further illustrate the same cytoplasmic flow field in a different  cross-section, now  perpendicular to the longitudinal axis and through the cylindrical region. The arrows illustrating in-plane flow in this cross-section (see horizontal color bar) now emphasize the azimuthal flows resulting from the helical actin organization, and the out-of-plane color map shows the reverse fountain streaming component. These results again illustrate the predictive power of our analytical modeling approach, including for the case of three-dimensional helical flows.

	\section{Discussion}\label{sec:discussion}
	
	\subsection{Summary} 
	In this paper, we have {used} lubrication theory  to derive a general solution for boundary-driven cytoplasmic flows in elongated cells.   We have then applied this framework to predict cytoplasmic fluid dynamics (and, when relevant, cortical stresses) in  {four} biologically relevant systems: the \textit{Drosophila} and \textit{C.~elegans} embryos (including pseudocleavage furrow formation), the pollen tube of seed plants, {and plant root hair cells}.  	 {Provided the flows are axisymmetric, our analytical lubrication solution is applicable even when an azimuthal component is present, and our model therefore comprehensively addresses axisymmetric flows in elongated cells.} 		By comparing our results with  experiments and numerical simulations, 
	we have shown that lubrication theory is a powerful tool to characterize cytoplasmic streaming; 
	our results therefore showcase the elegance and accuracy of fundamental  asymptotic solutions of the equations of fluid mechanics  in capturing   complex flows and stress patterns in a biological context.

	\subsection{Modeling limitations}
	Although lubrication theory is strictly only valid asymptotically in the limit of highly elongated cells, we obtain {surprisingly} accurate results even for modest aspect ratios. Lubrication theory is known, {empirically},  to work {unexpectedly} well outside its formal domain of applicability, and we see here that the case of boundary-driven cytoplasmic flows is no exception. To rationalize this, we note that classical lubrication theory is the leading order truncation of an asymptotic expansion in which the next term is second order in the inverse aspect ratio; in other words, our model incurs a quadratic error.
	
	This points to the obvious limitation of our approach: the model requires the  {wall velocity} $U(z,t)$  {and the geometry} $R(z)$, or more precisely, their second (spatial) derivatives, to vary on length scales significantly larger than the radial length scale.  While this is  a formal mathematical requirement, it can be relaxed in practice to some extent, as we demonstrated above.  {Clearly, this cannot be pushed too far,} and for instance, we cannot accurately model flows inside spherical cells with lubrication theory. 
	On the other hand, other well-established mathematical methods are available for solving Stokes' equations in spherical geometries~\cite{kimbook}.  Although beyond the scope of the current paper, higher-order terms in the lubrication expansion of Stokes' equations~\cite{tavakol2017} provide a promising avenue to improve the errors stemming from the requirement of slowly varying $U(z,t)$ and $R(z)$.
	
	Another strong assumption  of the models outlined here    is the requirement of axisymmetry (i.e.~no azimuthal dependence), which allows the work to be  analytically tractable by effectively reducing the number of dimensions by one. The model could be generalized to non-axisymmetric flows, but at the cost of more cumbersome algebra, with flows in each cross-section having to be solved in the form of infinite series-like solutions.

	\subsection{Outlook}

	Although we have focused on flows driven by tangential velocities along fixed walls, extensions of this fundamental framework point to exciting directions. For instance, our framework may be generalized to incorporate moving walls and wall-normal velocity boundary conditions. Indeed, lubrication solutions have been used to investigate peristaltic flows in the lumen of slime moulds~\cite{pringle2013} and the hydrodynamics inside actively contracting endoplasmic reticulum tubules~\cite{htet2024}. A more general long-wavelength solution with boundary conditions accounting for both tangential and normal wall motions could be used to investigate more dynamically complex problems, such as embryos in the early stages of cell division~\cite{hird1993}.    	
	Our approach may also be generalized to incorporate coupling between the cytoplasm and force generators at the cell boundary, as has been done computationally, for instance, in a two-fluid model of cytoplasmic streaming in the \textit{Drosophila} embryo~\cite{lopez2023} and with active gel theories of the cell cortex~\cite{bacher2019, bacher2021}. We believe our model offers a fundamental framework to build on and investigate the biomechanics of these active boundary forcings.

	{In this paper, we have illustrated that marrying even fundamental fluid mechanics with problems in developmental biology and plant science reveals a broad range of applications. We have developed an analytical model to accurately characterize fluid flows, a fundamental component of the various transport processes ubiquitous in biology, and our methods are therefore of interest to physicists and biologists alike investigating such transport phenomena in diverse biological settings. We are optimistic that the simplicity and accuracy of our model make it easily applicable beyond the four model systems we have investigated, and we thus hope to further inspire collaboration between biologists and modelers, and experimentalists and theoreticians, in the broad field of biophysics and biomechanics.}

	\section*{Acknowledgements}
	
	Financial support from the Cambridge Trust (scholarship to P.H.H.) is gratefully acknowledged.

	\section*{Data Availability}
	Modelling code is available in the Zenodo repository \url{https://doi.org/10.5281/zenodo.14861849} .

	\bibliography{AnalyticalMethodsForCS_12Feb2025.bib}
	
\end{document}